\documentclass[a4paper,12pt]{article}
\usepackage{amsmath,amsfonts,amssymb}
\usepackage{cite}

\begin{document}

\renewcommand*{\thefootnote}{\fnsymbol{footnote}}

\begin{center}
{\Large\bf Hydrogen-like symmetry in Regge spectrum of light mesons: selection of states}
\end{center}

\begin{center}
{S.S. Afonin\(^{a,b}\), 
A.V. Sarantsev\(^{b}\), A.M. Tsymbal\(^{a,b}\)
}
\end{center}

\begin{center}
  {\small\({}^a\)Saint Petersburg State University, 
  St.Petersburg 199034 Russia}\\
  \vspace*{0.15cm}
  {\small\({}^b\) NRC "Kurchatov Institute" -- PNPI, Gatchina 188300 Russia
  }
\end{center}

\renewcommand*{\thefootnote}{\arabic{footnote}}
\setcounter{footnote}{0}

\begin{abstract}
We discuss the $(L,n)$-classification of excited light non-strange mesons, where $L$ and $n$ are orbital and radial quantum numbers. The selection of true non-strange quark-antiquark excited states and assigning to them
definite $L$ and $n$ is a notoriously confusing problem. Three guiding principles for selection of correct observed states are formulated. They are applied for construction of a new $(L,n)$-classification. This classification
is consistent both with the approximate Regge form of the spectrum and with the hydrogen-like degeneracy, i.e., the dependence of mass on the sum $L+n$.
\end{abstract}

\bigskip

The study of spectrum of light mesons has a long history covered in many reviews~\cite{pdg}. The observed regularities in this spectrum encode important information on the nature of non-perturbative strong interactions.

As was first observed by Chew and Frautschi, the angularly excited light mesons form approximately linear Regge trajectories. Later it was observed that daughter Regge trajectories are almost equidistant, i.e., the radial meson trajectories are approximately linear as well~\cite{ani}. The analysis of experimental data on light non-strange mesons showed that the slopes of angular and radial trajectories are very close~\cite{bugg}. These observations
were later developed in a series of papers~\cite{cl1b,cl2,cl2b,cl3,klempt,shif}, where it was shown that the Regge spectrum of these mesons possesses the hydrogen-like degeneracy, $M^2(L,n)\sim L+n$, where $L$ and $n$ are orbital and radial quantum numbers. Recently this conclusion was confirmed in~\cite{Afonin:2024egd} using the modern data. The analysis of Ref.~\cite{Afonin:2024egd} was further discussed in~\cite{Afonin:2025yfx}.
The conservative Regge fit of the spectrum of light non-strange mesons performed in Ref.~\cite{Afonin:2025yfx} lead to the following parametrization of combined angular and radial trajectories (in GeV$^2$),
\begin{equation}
\label{fit}
\bar{M}^2 \approx (1.12\pm0.03)(L+n+0.54\pm0.09), \qquad L,n=0,1,2,3,4,
\end{equation}
where the ground states below 1~GeV, corresponding to $L+n=0$, were excluded.

The procedure of selection of light meson resonances used in the fit~\eqref{fit} was not explained in detail in Refs.~\cite{Afonin:2024egd,Afonin:2025yfx} neither in earlier similar fits.
The purpose of this note is to give a clear explanation for the choice of states in the $(L,n)$-classification of meson resonances.

The isoscalar light mesons practically always contain some admixture of $s\bar{s}$ component. The first important problem is that we need to separate out the mesons in which the strange component
dominates and not use them in the fit. Obviously, the $\phi$-mesons must be omitted. The case of other light mesons above 1~GeV is not so obvious.
The typical excitation energy in light mesons is 300--500~MeV, which is more than twice the mass of the $s$-quark.
In quantum field theory, this means that a part of the excitation energy can exist in the form of $s\bar{s}$-pair. This effect blurs the clear distinction between strange and non-strange excited light mesons.
The admixture of $s\bar{s}$ leads to the presence of $K\bar{K}$, $\eta$, $\eta'$ or $\phi$ in final states. But the presence of $s\bar{s}$ component does not necessarily mean
that it was a part of wave function of a given meson --- $s\bar{s}$-pair may simply have formed out of the gluon field, thereby causing the strong decay of that meson.
Moreover, decays via such a formation of $s\bar{s}$-pair may predominate because the $s$-quark is much heavier than the $u$ and $d$ quarks, hence it will be slower for a given energy, which is more favorable for the production of final stable hadrons. For example,  the letter "$\eta$" in the excited part of the spectrum, in reality, just denotes the isosinglet pseudoscalar particle --- the given notation does not mean that the resonance under consideration has the same quark composition as the ground state with these quantum numbers. The strange component is dominant in the $\eta$-meson (it gives the dominant contribution to the mass) but the same cannot be claimed for the high radial and orbital excitations of the $\eta$-meson. In particular, such excitations were observed in proton-antiproton annihilation~\cite{pdg,ani} where there is almost no $s\bar{s}$ component in the initial state. On the other hand, the formation of two $s\bar{s}$-pairs in the process of decay is much less probable. For this reason, if decay products of some meson contain a large fraction of $\eta\eta$, $\eta'\eta'$, $\phi\phi$ or $K\bar{K}K\bar{K}$ in its final states, we interpret this meson as predominantly an $s\bar{s}$ one. This will be our guiding principle. As mixing between flavor octet and singlet is usually close to the ideal in the excited light mesons, the resonances above 1~GeV with dominant $s\bar{s}$ component are flavor singlets. They are 200--300~MeV heavier than the corresponding flavor octet isosinglet resonances, we will call them "$s$-partners" of the latter. These $s$-partners must be excluded from our fit.

Next important problem is that in our $(L,n)$-classification many $(i,j)$ and $(j,i)$ states (i.e., states with equal sum $L+n$) have identical quantum numbers, the question arises which $L$ and $n$ should be assigned. For instance, there are vector mesons with
$L=0$ and $L=2$. We will follow the principle that the orbital excitations are easier to create: From the viewpoint of quantum mechanics, the excited state obtained by flipping the total quark spin is simpler (has a simpler wave function) than the radially excited state. For example, the $\rho$-meson is a $(0,0)$-state. Its second excitation, the $\rho(1700)$-meson, can be classified as $(2,0)$ (a $D$-wave state) or as $(0,2)$ (an $S$-wave state).
Following our principle, we must prefer the first possibility. If two nearly degenerate resonances with identical quantum numbers are observed, the one which is better seen (better established) should have larger $L$. If only
one resonance was seen at a predicted energy level (as with $\rho(1700)$), the largest possible $L$ should be assigned. This will be our second guiding principle. The given principle has a well-known empirical manifestation: The states lying on the main Regge trajectories are usually much better established than the states belonging to the corresponding daughter Regge trajectories.

The two principles formulated above are not enough --- we should also exclude suspicious resonances which highly likely are not genuine quark-antiquark states. The case of exotic states $\pi_1$ and $\eta_1$ is obvious.
The natural decay width of excited quark-antiquark states constitutes approximately 10-15\% of their mass~\cite{cl1b}. If a decay width notably deviates from this rule, the corresponding light meson, most likely, represents
a hybrid state or was erroneously determined as a resonance. This will be our third guiding principle.

Table~1 shows the $(L,n)$-classification of light non-strange mesons constructed by applying these three guiding principles. Excluded states above 1~GeV from the main list of PDG~\cite{pdg} and its section "Further States" are briefly commented in Tables~2 and~3, respectively. States denoted as $X$ were of course not included in any fit or analysis due to obvious reasons. 

We should note that the $(L,n)$-classification in Table~1 is somewhat different from the corresponding classification of previous papers on this subject including the recent Ref.~\cite{Afonin:2024egd}. The reason is that the three guiding principles formulated above were not completely taken into account in the previous papers.
It is therefore necessary to make new fits. Using all resonances from Table~1 we obtained the following estimate on the joined angular and radial trajectories,
\begin{equation}
\label{2}
\bar{M}^2_\text{all} = (1.15\pm0.03)L+(1.20\pm0.05)n+0.52\pm0.04.
\end{equation}
It is seen that the slopes of angular and radial trajectories are very close and can coincide within the experimental errors. The well motivated hypothesis of universal slope
leads to the fit
\begin{equation}
\label{3}
\bar{M}^2_\text{all} \approx  (1.16\pm0.03)(L+n+0.43\pm0.05).
\end{equation}
If we use only the states from the main list of PDG~\cite{pdg}, the fits~\eqref{2} and~\eqref{3} change to
\begin{equation}
\bar{M}^2_\text{main} = (1.14\pm0.07)L+(1.2\pm0.1)n+0.52\pm0.07,
\end{equation}
and
\begin{equation}
\bar{M}^2_\text{main} \approx  (1.15\pm0.06)(L+n+0.45\pm0.07),
\end{equation}
correspondingly. In any case, the new fits are close to the previous conservative estimate~\eqref{fit}.

\begin{table}
\vspace{-1cm}
\caption{
\small The $(L,n)$-classification of light non-strange mesons proposed in this work.}
\begin{center}
\begin{tabular}{|c|c|c|c|c|c|}
\hline
\begin{tabular}{c}
\begin{picture}(15,15)
\put(0,12){\line(1,-1){15}}
\put(-2,-3){$L$}
\put(10,7){$n$}
\end{picture}\\
\end{tabular}
& 0 & 1 & 2 & 3 & 4 \\
\hline
0
&
\begin{tabular}{c}
$\pi$\\
$\rho$\\
$\omega$\\
\end{tabular}
&
\begin{tabular}{c}
$\pi(1300)$\\
$\eta(1295)$\\
$\rho(1450)$ \\
$\omega(1420)$ \\
\end{tabular}
&
\begin{tabular}{c}
$\pi(1800)$\\
$\eta(1760)$\\
$\rho(?)$\\
$\omega(?)$\\
\end{tabular}
&
\begin{tabular}{c}
$\pi(2070)$\\
$\eta(2010)$\\
$\rho(?)$ \\
$\omega(?)$\\
\end{tabular}
&
\begin{tabular}{c}
$\pi(2360)$\\
$\eta(2320)$\\
$\rho(?)$\\
$\omega(?)$ \\
\end{tabular}
\\
\hline
1
&
\begin{tabular}{c}
$a_0(1450)$\\
$f_0(1370)$\\
$a_1(1260)$\\
$f_1(1285)$\\
$b_1(1235)$\\
$h_1(1170)$\\
$a_2(1320)$\\
$f_2(1270)$\\
\end{tabular}
&
\begin{tabular}{c}
$a_0(1710)$\\
$f_0(1710)$\\
$a_1(1640)$\\
$f_1(?)$\\
$b_1(?)$ \\
$h_1(?)$ \\
$a_2(1700)$\\
$f_2(1750)$\\
\end{tabular}
&
\begin{tabular}{c}
$a_0(2020)$\\
$f_0(2020)$\\
$a_1(1930)$ \\
$f_1(1970)$\\
$b_1(1960)$\\
$h_1(1965)$\\
$a_2(2030)$ \\
$f_2(2000)$\\
\end{tabular}
&
\begin{tabular}{c}
$a_0(?)$\\
$f_0(2200)$\\
$a_1(2270)$ \\
$f_1(2310)$\\
$b_1(2240)$\\
$h_1(2215)$\\
$a_2(2175)$ \\ 
$f_2(2295)$\\
\end{tabular}
&\\
\hline
2
&
\begin{tabular}{c}
$\rho(1700)$\\
$\omega(1650)$\\
$\pi_2(1670)$\\
$\eta_2(1645)$\\
$\rho_2(?)$\\
$\omega_2(?)$\\
$\rho_3(1690)$\\
$\omega_3(1670)$\\
\end{tabular}
&
\begin{tabular}{c}
$\rho(2000)$\\
$\omega(1960)$\\
$\pi_2(2005)$\\
$\eta_2(2030)$\\
$\rho_2(1940)$\\
$\omega_2(1975)$\\
$\rho_3(1990)$\\
$\omega_3(1945)$\\
\end{tabular}
&
\begin{tabular}{c}
$\rho(2270)$\\
$\omega(2290)$ \\
$\pi_2(2285)$\\
$\eta_2(2250)$\\
$\rho_2(2225)$\\
$\omega_2(2195)$\\
$\rho_3(?)$ \\
$\omega_3(2285)$\\
\end{tabular}
&  &\\
\hline
3
&
\begin{tabular}{c}
$a_2(1990)$\\
$f_2(1950)$\\
$a_3(2030)$\\
$f_3(2050)$\\
$b_3(2030)$\\
$h_3(2025)$\\
$a_4(1970)$\\
$f_4(2050)$\\
\end{tabular}
&
\begin{tabular}{c}
$a_2(2255)$ \\
$f_2(2240)$\\
$a_3(2275)$\\
$f_3(2300)$\\
$b_3(2245)$\\
$h_3(2275)$\\
$a_4(2255)$\\
$f_4(2300)$\\
\end{tabular}
&  &  &\\
\hline
4
&
\begin{tabular}{c}
$\rho_3(2250)$\\
$\omega_3(2255)$\\
$\pi_4(2250)$\\
$\eta_4(2330)$\\
$\rho_4(2230)$\\
$\omega_4(2250)$ \\
$\rho_5(2350)$\\
$\omega_5(2250)$\\
\end{tabular}
&  &  &  &\\
\hline
\end{tabular}
\end{center}
\end{table}

\begin{table}
\vspace{-1cm}
\caption{
\small Excluded light mesons above 1~GeV from the main list of PDG~\cite{pdg}, see other comments on these states in~\cite{pdg}.}
\vspace{-0.5cm}
\begin{center}
\begin{tabular}{|c|l|}
\hline
\small Resonance & \hspace{3.9cm} \small Reasons of exclusion\\
\hline
$\pi_1(1400)$ & \small Exotic state in the quark model\\
$\eta(1405)$ & \small Considered a manifestation of the same state as $\eta(1475)$~\cite{pdg}\\
$h_1(1415)$ & \small $s$-partner of $h_1(1170)$, also unusually small width $78\pm11$ MeV\\
$f_1(1420)$ & \small May represent $f_1(1510)$ seen in a different channel~\cite{pdg}\\
$f_2(1430)$ & \small Unusually small width (10--80~MeV)\\
$\eta(1475)$ & \small $s$-partner of $\eta(1295)$\\
$f_0(1500)$ & \small Likely it is the radially excited $f_0(980)$ in which the $s$ component\\
& \small dominates~\cite{Sarantsev:2021ein}; possibly contains a large admixture of glueball~\cite{pdg}\\
$f_1(1510)$ & \small $s$-partner of $f_1(1285)$\\
$f_2'(1525)$ & \small $s$-partner of $f_2(1270)$\\
$f_2(1565)$ & \small There are many doubts that it is a true pole~\cite{pdg}\\
$\rho(1570)$ & \small It is a $\phi\pi$ state; may be an OZI-violating decay mode of $\rho(1700)$~\cite{pdg}\\
$h_1(1595)$ & \small Unusually large width (about 400~MeV~\cite{pdg}); observed only in one old\\
& \small experiment, not seen now by BES3\\
$\pi_1(1600)$ & \small Exotic state in the quark model\\
$f_2(1640)$ & \small Extra state for our classification; may coincide with $f_2(1750)$\\
$f_0(1770)$ & \small Replaced by $f_0(1710)$ which naturally lies on the radial trajectory \\
& \small containing $f_0(1370)$~\cite{Sarantsev:2021ein}; seen mainly in $\phi\omega$ mode~\cite{pdg,Sarantsev:2021ein}\\
$f_2(1810)$ & \small Often contains $\eta\eta$ in its decay products\\
$\eta_1(1855)$ & \small Exotic state in the quark model\\
$\eta_2(1870)$ & \small The $s$-partner of $\eta_2(1645)$\\
$\pi_2(1880)$ & \small Replaced by $\pi_2(2005)$; candidate for a hybrid state~\cite{Anisovich:2001hj}\\
$\rho(1900)$ & \small Most likely, emerges as a $N\bar{N}$ threshold effect~\cite{pdg}\\
$f_2(1910)$ & \small Replaced by $f_2(1950)$ seen in a much larger number of reactions; \\
& there is an $\eta\eta'$ mode\\
$a_0(1950)$ & \small Seems to coincide with $a_0(2020)$\\
$f_2(2010)$ & \small $\phi\phi$ mode is dominant, likely is the $s$-partner of $f_2(1750)$\\
$\pi_2(2100)$ & \small Unusually large width ($620\pm50$~MeV~\cite{pdg})\\
$f_0(2100)$ & \small Often $\eta\eta$ in its final states; naturally belongs to the radial trajectory \\
& \small containing $f_0(980)$, $f_0(1500)$ and $f_0(1770)$~\cite{Sarantsev:2021ein} \\
$f_2(2150)$ & \small $\eta\eta$ and $\phi\phi$ in its final states\\
$\rho(2150)$ & \small Its mass value is highly controversial, PDG~\cite{pdg} does not report any \\
& averaged one\\
$\omega(2220)$ & \small Unusually small width ($105\pm34$~MeV~\cite{pdg}); replaced by $\omega(2290)$ \\
$\eta(2225)$ & \small $4K$ in its decay products; likely is the $s$-partner of $\eta(2010)$\\
$f_2(2300)$ & \small $\phi\phi$ in its final states; replaced by $f_2(2295)$\\
$f_0(2330)$ & \small Often $\eta\eta$ and $\eta'\eta'$ in its final states; naturally lies on the radial \\
& \small trajectory beginning with $f_0(980)$ --- next state after $f_0(2100)$~\cite{Sarantsev:2021ein} \\
$f_2(2340)$ & \small  Often $\eta\eta$, $\eta'\eta'$ and $\phi\phi$ in its decay products\\
$f_0(2470)$ & \small Unusually small width (about $75$~MeV~\cite{pdg}); decays into $\eta'\eta'$ \\
$f_6(2510)$ & \small Looks like a $L=5$ state --- our fits include only $L\leq4$ states\\
\hline
\end{tabular}
\end{center}
\end{table}

\begin{table}
\vspace{-1cm}
\caption{
\small Excluded light mesons above 1~GeV from the section "Further States" of PDG~\cite{pdg}.}
\vspace{-0.5cm}
\begin{center}
\begin{tabular}{|c|l|}
\hline
\small Resonance & \hspace{3.9cm} \small Reasons of exclusion\\
\hline
$a_3(1875)$ & \small Unusually large width (about $385$ MeV)\\
$h_1(1900)$ & \small Is probably the same state as $h_1(1965)$\\
$a_2(1950)$ & \small Observed on the edge of energy measured in the Crystall Barrell experiment,\\
& \small  is therefore very speculative\\
$\pi_1(2015)$ & \small An exotic state in the quark model\\
$f_0(2060)$ & \small Its mass is not well defined, PDG provides $\sim2050$ MeV \\
$a_1(2095)$ & \small Unusually large width (around $450$ MeV) \\
$\eta(2100)$ & \small 4K in its decay products, may coincide with $\eta(2225)$ \\
$\omega(2120)$ & \small It may also be $\phi(2120)$, and its spin is not defined~\cite{pdg} \\
$f_2(2140)$ & \small Unusually small width ($49\pm28$ MeV) \\
$\eta(2190)$ & \small Unusually large width ($850\pm100$ MeV) \\
$\omega(2330)$ & \small Unusually large width ($435\pm75$ MeV), replaced by $\omega(2290)$ \\
$a_6(2450)$ & \small Should be a $L=5$ state and our fit includes only $L\leq4$ states; \\
& \small may be a partner of $f_6(2510)$ from the previous table \\
$f_6(3100)$ & \small Too heavy (looks like a $L=7$ state) \\
\hline
\end{tabular}
\end{center}
\end{table}

\underline{\bf Acknowledgements}. The paper is supported by the RNF grant 24-22-00322.


\end{document}